\documentclass[a4paper]{article}
\usepackage[utf8]{inputenc}
\usepackage[T1]{polski}
\usepackage{graphicx}\graphicspath{{img/}}
\usepackage{lmodern,fancyhdr,secdot,float,amsmath,amsfonts,amsthm,amssymb}
\usepackage{booktabs,tabularx}
\usepackage[margin=2cm]{geometry}
\usepackage{multicol}
\usepackage[hidelinks,unicode]{hyperref}
\usepackage{titlesec}
\usepackage{multirow}
\usepackage{tabularx}
\usepackage[table,xcdraw]{xcolor}
\titleformat{\section}{\centering\normalsize\bf}{\thesection.}{.5em}{\MakeUppercase}
\titleformat*{\subsection}{\bf\normalsize\selectfont}
\titleformat*{\subsubsection}{\bf\normalsize\selectfont}
\newcommand{\titlePL}[1]{\large\textbf{ #1}}
\newcommand{\titleEN}[1]{\normalsize #1}
\newcommand{\keywordsPL}[1]{\small\textbf{Słowa kluczowe:} #1}
\newcommand{\keywordsEN}[1]{\small\textbf{Keywords:} #1}
\newcommand{\abstractPL}[1]{\small\textbf{Streszczenie:} #1}
\newcommand{\abstractEN}[1]{\small\textbf{Abstract:} #1}

\begin{document}\thispagestyle{empty}\pagestyle{fancy}
\begin{minipage}[t]{0.5\textwidth}\vspace{0pt}%
\includegraphics[scale=1.1]{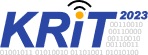}
\end{minipage}
\begin{minipage}[t]{0.45\textwidth}\vspace{0pt}%
\centering
KONFERENCJA RADIOKOMUNIKACJI\\ I TELEINFORMATYKI\\ KRiT 2023
\end{minipage}

\vspace{1cm}

\begin{center}
\titlePL{Systemy komórkowe 5G wspierane przez BSP, OZE oraz IPR}

\titleEN{5G cellular systems supported by UAVs, RESs, and RISs}\medskip

Adam Samorzewski$^{1}$;
Adrian Kliks$^{2,3}$; 

\medskip

\begin{minipage}[t]{1\textwidth}
\centering
\small $^{1}$ Politechnika Poznańska, Poznań, Polska, \href{mailto:email}{adam.samorzewski@doctorate.put.poznan.pl}\\
\small $^{2}$ Politechnika Poznańska, Poznań, Polska, \href{mailto:email}{adrian.kliks@put.poznan.pl}\\
\small $^{3}$ Uniwersytet Techniczny w Luleå, Luleå, Szwecja, \href{mailto:email}{adrian.kliks@ltu.se}\\
\end{minipage}

\medskip
\end{center}

\medskip

\begin{multicols}{2}
\noindent
\abstractPL{
W artykule rozważono zużycie energetyczne w sieciach komórkowych 5G zasilanych przez Odnawialne Źródła Energii (OZE) oraz wyposażonych w~Inteligentne Powierzchnie Rekonfigurowalne (IPR) oraz Bezzałogowe Statki Powietrzne (BSP) \textit{na uwięzi}, pełniące rolę mobilnych punktów dostępowych. Badania skoncentrowano na energetycznej stronie sieci dostępu radiowego (ang. Radio Access Network -- RAN) zlokalizowanej na terenie miasta Poznania w Polsce. Zysk związany z~wykorzystaniem generatorów OZE, czyli paneli fotowoltaicznych (ang. photovoltaic panels -- PVP) dla stacji bazowych (ang. base stations -- BS) został przedstawiony w~postaci dwóch współczynników: średniej liczby ładowań BSP (SLLB), aby zapewnić ciągły dostęp do usług mobilnych dla obsługiwanych terminali mobilnych użytkowników (ang. user equipment -- UE) oraz średniej redukcji zużycia energii (SRZE) przez system bezprzewodowy.}
\medskip

\noindent
\abstractEN{
The paper considers energy consumption in 5G cellular networks powered by Renewable Energy Sources (RESs) and equipped with Reconfigurable Intelligent Surfaces (RISs) and tethered Unmanned Aerial Vehicles (UAVs), acting as mobile access points. The study was focused on the energy side of the Radio Access Network (RAN) located in the city of Poznań in Poland. The profit associated with the use of renewable energy generators, i.e. photovoltaic panels (PVP) for base stations (BSs) is presented in the form of two factors: the average number of UAV charges (ANUC) to provide continuous access to mobile services for connected user equipment (UE) terminals, and the average reduction in energy consumption (AREC) of the wireless system.\footnote{Copyright © 2023 SIGMA-NOT. Personal use is permitted. For any other purposes, permission must be obtained from the SIGMA-NOT by emailing sekretariat@sigma-not.pl. This is the author’s version of an article that has been published in the journal entitled \textit{Telecommunication Review -- Telecommunication News} (PL: \textit{Przegląd Telekomunikacyjny -- Wiadomości Telekomunikacyjne}) by the SIGMA-NOT. Changes were made to this version by the publisher before publication, the final version of the record is available at: https://dx.doi.org/10.15199/59.2023.4.18. To cite the paper use: A. Samorzewski, A.~Kliks, “5G cellular systems supported by UAVs, RESs, and RISs" (PL: "Systemy komórkowe 5G wspierane przez BSP, OZE oraz IPR"), \textit{Telecommunication Review -- Telecommunication News} (PL: \textit{Przegląd Telekomunikacyjny -- Wiadomości Telekomunikacyjne}), 2023, no. 4, pp.~97--100, doi: 10.15199/59.2023.4.18 or visit https://sigma-not.pl/publikacja-144746-2023-4.html.}}
\medskip

\noindent
\keywordsPL{5G, Bezzałogowe Statki Powietrzne, Inteligentne Powierzchnie Rekonfigurowalne, Odnawialne Źródła Energii, sieci komórkowe}
\medskip

\noindent
\keywordsEN{5G, cellular networks, Reconfigurable Intelligent Surfaces, Renewable Energy Sources, Unmanned Aerial Vehicles}

\section{Wstęp}
Obecne systemy telekomunikacyjne zasilane są głównie przez konwencjonalne źródła energii (KZE), tj., paliwami kopalnymi. Systemy te odpowiadają za około $25\%$ ilości emisji dwutlenku węgla ($\text{CO}_2$) powodowanych przez segment ICT (ang. \textit{Information and Communication Technology}), którego zapotrzebowanie na energię wydaje się być coraz większe z roku na rok~\cite{I}.
Ponadto sektor ICT jest odpowiedzialny za sporą część globalnych emisji gazów cieplarnianych (ang. \textit{Greenhouse Gas -- GHG}), której możliwie niedoszacowana wartość może sięgać nawet od $2,1$ do $3,9\%$~\cite{Freitag}. Zatem, aby zredukować wspomnianą ilość gazów, konieczne może okazać się znalezienie alternatywnych źródeł energii, które zarówno zaspokoją zapotrzebowanie na energię systemów bezprzewodowych ($4$G, $5$G oraz kolejnych generacji), jak i zmniejszą ilość $\text{CO}_2$ emitowanego do atmosfery. Wówczas zaangażowanie odnawialnych źródeł energii (OZE), np. paneli fotowoltaicznych, jest rozwiązaniem godnym uwagi ze względu na sposób pozyskiwania zasobów z promieniowania słonecznego, który sam w sobie wydaje się być mniej szkodliwy dla środowiska, tj., niezanieczyszczający. Co więcej, ilość potencjalnie generowanych zasobów energetycznych można by uznać w pewnym stopniu za niewyczerpalną (wyłączając konieczność wymiany paneli fotowoltaicznych po czasie ściśle określonym przez ich producenta) w kontekście długoterminowym~\cite{Deruyck, Abid}.

W ramach koncepcji systemów komórkowych $5$-ej generacji usługi mobilne dzielą się na cztery główne grupy, z~których trzy dotyczą odpowiednio wysokich przepływności, małych opóźnień oraz dużej liczby jednocześnie podłączonych urządzeń~\cite{ 3GPP:TR21.915}. Z punktu widzenia realizacji pierwszej z~nich, można napotkać ograniczenie związane ze skończoną liczbą bloków zasobów fizycznych (ang. \textit{Physical Resource Blocks} -- PRB) przypadającą na komórkę sieci. Problem ten można zauważyć zwłaszcza w obszarach miejskich, gdzie zaludnienie jest dość gęste. Zatem, aby zapewnić wystarczającą pojemność sieci bezprzewodowych na tego typu obszarach, zaproponowano rozwiązanie w~postaci stacji bazowych obejmujących regiony tzw. małymi komórkami, które opiera się na rozmieszczaniu węzłów dostępowych małej mocy bardzo blisko siebie. Jednak przy wdrażaniu tej koncepcji ze względu na utrudnienia, takie jak niekorzystna architektura miejska czy braki finansowe na budowę nowych stacjonarnych stacji bazowych, istnieje ryzyko pojawienia się \textit{luki sygnałowej}. Jednym ze sposobów uniknięcia tego zjawiska, coraz częściej branym pod uwagę w pracach naukowych, jest zaangażowanie dodatkowego sprzętu, np. bezzałogowych statków powietrznych (BSP) jako mobilnych stacji bazowych (ang. \textit{Mobile Base Station} -- MBS). To z kolei daje operatorom sieci komórkowych (ang. \textit{Mobile Network Operator} -- MNO) możliwość dynamicznego dostosowywania aktualnej lokalizacji węzłów dostępowych w celu pokrycia terenów, do których sygnał radiowy nie dociera, i/lub obsługi istniejącej infrastruktury telekomunikacyjnej w obszarach miejskich, gdzie liczba jednocześnie podłączonych terminali użytkowników może ulegać wahaniom, a nawet przekraczać pierwotnie założoną pojemność systemu (np. z powodu wydarzeń publicznych)~\cite{Mozaffari, Alzenad}.

W dodatku wykorzystanie tak zwanych inteligentnych powierzchni rekonfigurowalnych (IPR) do sterowania pokryciem radiowym systemów bezprzewodowych cieszy się dużym zainteresowaniem w aktualnej literaturze. IPR to urządzenie w postaci powierzchni składającej się z dużej liczby pasywnych (lub czasami aktywnych) elementów odbijających, które są w stanie samodzielnie wywołać pożądaną zmianę fazy i/lub amplitudy padającego sygnału radiowego. Dzięki zapewnieniu elastycznej rekonfiguracji propagacji sygnału (po włączeniu IPR), operatorzy mobilni byliby w stanie osiągnąć lepszą wydajność swoich sieci poprzez zmniejszenie zakłóceń i przerw oraz podniesienie niezawodności, pojemności i przepływności łączy radiowych~\cite{Huang, Di Renzo}.

W niniejszym artykule zbadano wydajność urządzeń BSP \textit{na uwięzi} (przyłączonych przewodem do zewnętrznego zestawu bateryjnego umiejscowionego na ziemi/budynku) pracujących jako stacje bazowe, które są wyposażone zarówno w IPR (do potencjalnego wykorzystania w przyszłości), jak i OZE (do wydłużenia czasu pracy). Celem tej analizy było znalezienie najodpowiedniejszego rozmieszczenia i konfiguracji mobilnych stacji bazowych na określonym obszarze przy uwzględnieniu mocy nadawania, energii zużywanej przez BSP oraz produkcji energii przez OZE. Kolejno oszacowano kompromis między dodatkową masą potrzebną do przenoszenia elementów OZE i IPR a korzyściami związanymi z~energią wytwarzaną przez OZE. Na koniec przeprowadzono symulacje opierając się na rzeczywistych danych atmosferycznych (zawarte w \cite{WeatherData}), aby zweryfikować faktyczną wydajność proponowanego rozwiązania.

\section{Scenariusz systemowy}
\begin{figure}[H]
\centering
\includegraphics[width=0.475\textwidth]{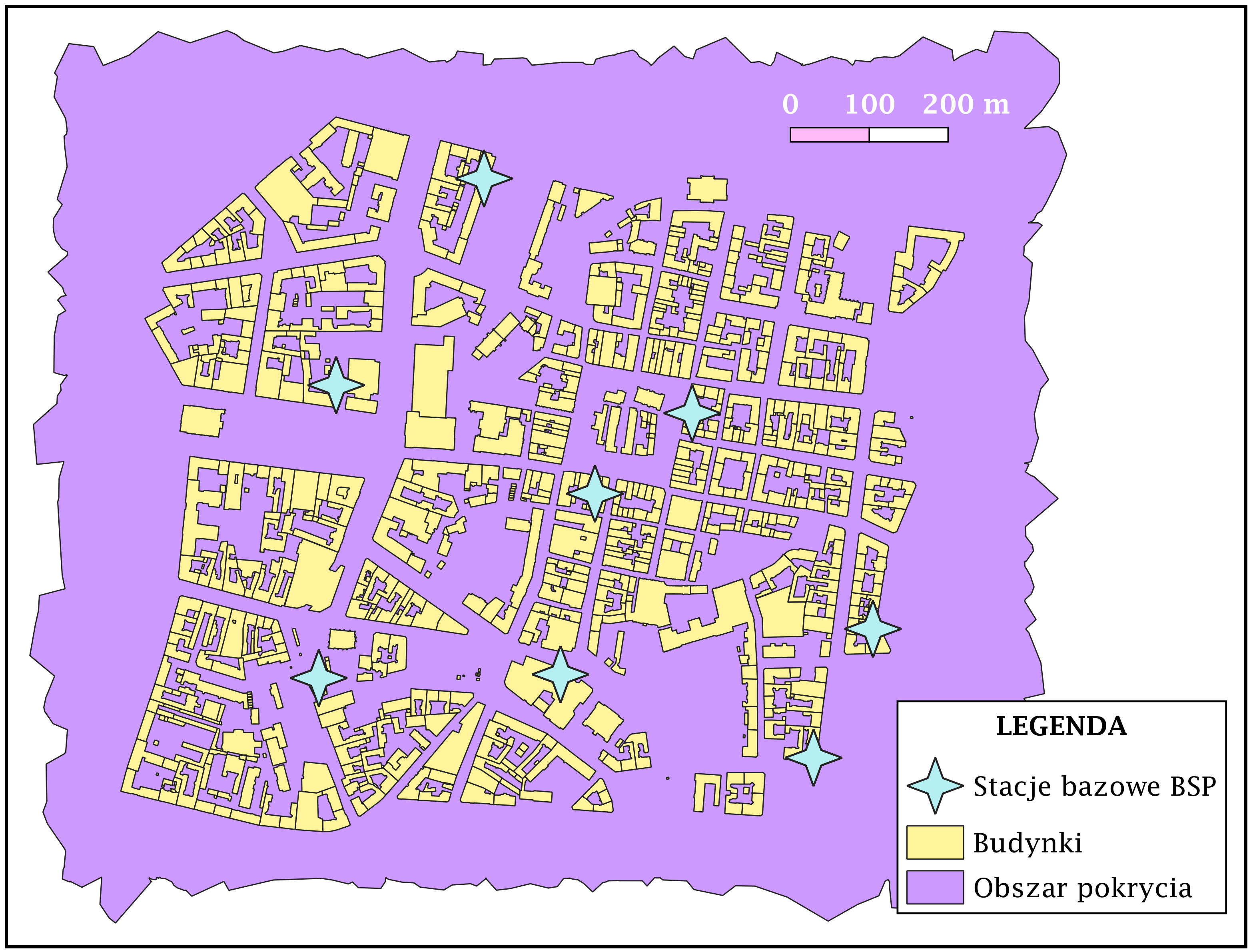}
\caption{Mapa badanego obszaru (Poznań).}
\label{figure:map}
\end{figure}

Rozważany w pracy scenariusz uwzględnia sieć komórkową $5$G zlokalizowaną w obrębie miasta Poznania w Polsce (dane z \cite{PoznanData}), której stacjami bazowymi są bezzałogowe statki powietrzne zlokalizowane na podstawie rzeczywistych danych jednego z polskich operatorów sieci komórkowych (zawarte w \cite{NetworkData}) i unoszące się $50$ m nad ziemią. W~przeprowadzonych badaniach założono, że rozmieszczenie dronów jest stacjonarne w czasie obsługi wspomnianego obszaru. Co więcej, na terenie sieci bezprzewodowej jest $100$ losowo rozmieszczonych użytkowników, gdzie każdy z nich wymaga stałej przepływności równej $100$ Mb/s na łączu w dół oraz $25$ Mb/s na łączu w górę. Na Rys.~\ref{figure:map} zamieszczono mapę przedstawiającą opisany powyżej scenariusz wdrożenia systemu bezprzewodowego.
\subsection{Projektowanie sieci komórkowej}
\label{subsection:design}
Rozważana sieć komórkowa została zaprojektowana przy użyciu specjalistycznego oprogramowania opisanego w \cite{Castellanos} o~nazwie Green Radio Access Network Design (GRAND). Wspomniane narzędzie optymalizuje również działanie sieci dostępu radiowego pod kątem zużycia energii i/lub narażenia ludzi na działanie pola elektromagnetycznego (PEM). Proces optymalizacyjny opiera się na wyznaczeniu optymalnej liczby włączonych komórek stacji bazowych (o predefiniowanych lokalizacjach) ze wszystkich dostępnych oraz dostosowanie ich parametrów transmisji sygnału radiowego w oparciu o chwilowe przepływności wymagane przez użytkowników podłączonych do systemu bezprzewodowego. Jako dane wejściowe oprogramowanie GRAND otrzymuje listy dostępnych stacji bazowych i aktywnych użytkowników oraz pliki $3$D opisujące rozważany obszar pokrycia oraz znajdujące się na jego terenie budowle. Lokalizacja każdego użytkownika jest wybierana losowo przez symulator w każdym pojedynczym przebiegu. Dodatkowo na samym początku wszystkie układy nadawczo-odbiorcze (NO) węzłów dostępowych nadają sygnał radiowy z maksymalną mocą. Po rozmieszczeniu użytkowników na badanym terenie program wybiera, które komórki sieci mają zostać włączone i z jaką mocą powinny przesyłać dane, aby dotrzeć do jak największej liczby aktywnych użytkowników a jednocześnie by zminimalizować ich narażenie na promieniowanie PEM i/lub zużycie energetyczne systemu. Na tym etapie narzędzie GRAND ocenia potencjalne połączenia radiowe między siecią a każdym UE w oparciu zarówno o ich wymagania transmisyjne, jak i również maksymalnie dopuszczalną wartość tłumienia propagowanego sygnału przy założeniu różnych konfiguracji mocy nadawania. Następnie, w celu znalezienia optymalnego rozwiązania, formułuje się funkcję celu opierając się na przewidywanej optymalizacji narażenia ludzi na działanie PEM, zużycia energii lub obu. Następnie wszyscy użytkownicy rozpoczynają wymianę danych z odpowiednimi węzłami dostępowymi w sposób ciągły, powodując stałe obciążenie ruchem przez cały czas trwania pojedynczej symulacji. Równolegle wykonywane są obliczenia bilansu energetycznego dla wszystkich stacji bazowych, które następnie przekładają się na średnią liczbę potrzebnych doładowań akumulatorów na UAV z i bez OZE.

\subsection{Wyposażenie}
\label{subsection:equipment}
Implementacja paneli fotowoltaicznych w ramach oprogramowania GRAND została zainspirowana specyfikacją rzeczywistego urządzenia, którą można znaleźć w \cite{PvData}. Przyjęto, że wykorzystane panele fotowoltaiczne są montowane na górnej części pokrywy UAV w postaci cienkowarstwowych ogniw słonecznych. W związku z tym, ze względu na znikomy wpływ na całkowity pobór mocy, ciężar paneli fotowoltaicznych został pominięty we wszystkich obliczeniach poświęconych charakterystykom energetycznym stacji BSP. Dodatkowo obliczenia te przeprowadzono dla czterech różnych dni (każdy rozpoczynający inną porę roku), aby pokazać, jak pozyskiwanie energii przez panele fotowoltaiczne zależy od pory roku dla klimatu panującego Polsce.

Podobnie jak w przypadku paneli fotowoltaicznych, systemy bateryjne dla stacji mobilnych BSP również zostały zaprojektowane na podstawie rzeczywistej implementacji opisanej w \cite{BatteryData}. Celem wyposażania dronów w akumulatory jest zaspokojenie ich zapotrzebowanie na energię, a także magazynowanie zasobów wytwarzanych przez generatory OZE. Rozładowanie ogniw baterii pociąga za sobą konieczność wymiany stacji BSP na w pełni naładowaną. Niemniej jednak przyjęto założenie, że akumulatory można całkowicie rozładować, a $5\%$ całkowitej energii zgromadzonej w jednym akumulatorze jest zawsze zużywane na loty urządzenia typu dron z i do stacji doładowującej.

Układy nadawczo-odbiorcze węzłów dostępowych BSP pracują zgodnie z technologią Multiple-Input-Multiple-Output (MIMO) wykorzystując $64$ aktywne elementy antenowe (ang. \textit{active antenna element} -- AAE) transmitujące sygnały radiowe w paśmie częstotliwości $3500$ MHz. Przyjęte w oprogramowaniu metody estymacji kanału i przetwarzania sygnału są zgodne ze schematem minimalnego błędu średnio-kwadratowego (ang. \textit{minimum mean-squared error} -- MMSE). Ponadto każda mobilna stacja bazowa posiada jedno urządzenie typu IPR z $16.$ identycznymi pasywnymi elementami odbijającymi, które realizują przesunięcie fazowe sygnału radiowego z rozdzielczością $6$-bitową. W niniejszej pracy pominięto wpływ działania matryc IPR na propagację sygnału radiowego na badanym obszarze, w przeciwieństwie do ich oddziaływania na bilans energetyczny systemu bezprzewodowego. Jednakże wpływ urządzeń typu IPR na wydajność sieci komórkowej zostanie wzięty pod uwagę w~przyszłych rozważaniach.

\subsection{Modele energetyczne}
\label{subsection:models}
Wszystkie modele energetyczne użyte do przeprowadzenia operacji matematycznych w GRAND (związanych zarówno z produkcją, jak i konsumpcją energii – prosumpcją) zostały bezpośrednio zaczerpnięte lub zainspirowane literaturą naukową lub rzeczywistymi wdrożeniami. Lista tych formuł została zawarta w Tab.~\ref{table:models}.

\begin{table}[H]
\centering
\caption{Lista modeli energetycznych}
\label{table:models}
\resizebox{0.48\textwidth}{!}{
\begin{tabular}{|l|c|c|}
\hline
\multicolumn{1}{|c|}{Nazwa modelu energetycznego}              & Oznaczenie            & Źródło  \\ \hline
Zużycie energii przez wielowirnikowy BSP          & $P_\text{BSP}$  & \cite{Janji} \\ \hline
Zużycie energii przez MIMO układ NO                   & $P_\text{MIMO}$ & \cite{Björnson} \\ \hline
Zużycie energii przez matrycę IPR                          & $P_\text{IPR}$  & \cite{Huang} \\ \hline
Moc wyjściowa panelu fotowoltaicznego                               & $P_\text{PVP}$   & \cite{HomerPro:v3.15} \\ \hline
Składownie energii w systemie bateryjnym                               & $P_\text{BAT}$   & \cite{Voltacon} \\ \hline
\end{tabular}}
\end{table}

\section{Konfiguracja symulacji}
\label{section:simulation}
Kod źródłowy opracowanego oprogramowania został przygotowany w języku Java. Badanie opisanego powyżej scenariusza systemowego zostało przeprowadzone w formie $10$ niezależnych przebiegów symulacyjnych, z których każdy uwzględnia $4$ dni poprzedniego roku rozpoczynające różne pory roku -- równonoc wiosenną $\big(20^\text{th}$ marca $2022\big)$, przesilenie letnie $\big(21^\text{st}$ czerwca $2022\big)$, równonoc jesienną $\big(23^\text {rd}$ września $2022\big)$ oraz przesilenie zimowe $\big(21^\text{st}$ grudnia $2022\big)$. Parametry użytkowników (koordynaty współrzędnych oraz żądane warunki transmisyjne) są definiowane zawsze na początku każdego przebiegu symulacyjnego. Przyjęty krok czasowy odpowiada $1$ minucie czasu symulacyjnego $\left(4 \text{ dni}\cdot24\text{ godziny}\cdot60\text{ minut}=5760 \right.$ kroków przypadających na pojedynczą symulację$\left.\right)$, w którym aktualizowane są dane dotyczące warunków pogodowych, a następnie wykonywane są obliczenia dotyczące generacji i zużycia energetycznego przez komponenty systemu bezprzewodowego. 
\begin{table}[H]
\centering
\caption{Bilans energetyczny dla BSP z i bez OZE}
\label{table:energy_characteristics}
\resizebox{0.48\textwidth}{!}{
\begin{tabular}{|c|cc|}
\hline
                                      & \multicolumn{2}{c|}{Całkowita (i szczytowa) energia z OZE $\left[\text{Wh}\right]$} \\ \cline{2-3}
\multirow{-2}{*}{}                    & \multicolumn{1}{c|}{\quad\quad\textit{Bez OZE}\quad\quad\quad}       & \multicolumn{1}{c|}{\textit{PVP}}         \\ \hline
Równonoc wiosenna                        & \multicolumn{1}{c|}{$0 \text{ } \left(0\right)$}        & $475,17 \text{ } \left(60,57\right)$         \\ \hline
Przesilenie letnie                       & \multicolumn{1}{c|}{$0 \text{ } \left(0\right)$}        & $572,64 \text{ } \left(91,86\right)$         \\ \hline
Równonoc jesienna                        & \multicolumn{1}{c|}{$0 \text{ } \left(0\right)$}        & $349,56 \text{ } \left(65,15\right)$         \\ \hline
Przesilenie zimowe                       & \multicolumn{1}{c|}{$0 \text{ } \left(0\right)$}        & $17,67 \text{ } \left(4,18\right)$         \\ \hline
\rowcolor[HTML]{6665CD} 
{\color[HTML]{FFFFFF} Wartość średnia} & \multicolumn{1}{c|}{\cellcolor[HTML]{6665CD}{\color[HTML]{FFFFFF} $0$}} & {\color[HTML]{FFFFFF} $353,76$} \\ \hline
\end{tabular}}

\begin{tabular}{cc}
     \\
\end{tabular}

\resizebox{0.48\textwidth}{!}{
\begin{tabular}{|c|cc|}
\hline
                                      & \multicolumn{2}{c|}{Średnia redukcja zużycia energii (SRZE) $\left[\text{\%}\right]$} \\ \cline{2-3}
\multirow{-2}{*}{}                    & \multicolumn{1}{c|}{\quad\quad\textit{Bez OZE}\quad\quad\quad}       & \multicolumn{1}{c|}{\textit{PVP}}         \\ \hline
Równonoc wiosenna                        & \multicolumn{1}{c|}{$0$}                                                & $7,28$                        \\ \hline
Przesilenie letnie                       & \multicolumn{1}{c|}{$0$}                                                & $8,71$                        \\ \hline
Równonoc jesienna                        & \multicolumn{1}{c|}{$0$}                                                & $5,33$                        \\ \hline
Przesilenie zimowe                       & \multicolumn{1}{c|}{$0$}                                                & $0,27$                        \\ \hline
\rowcolor[HTML]{6665CD} 
{\color[HTML]{FFFFFF} Wartość średnia} & \multicolumn{1}{c|}{\cellcolor[HTML]{6665CD}{\color[HTML]{FFFFFF} $0$}} & {\color[HTML]{FFFFFF} $5,4$} \\ \hline
\end{tabular}}

\begin{tabular}{cc}
     \\
\end{tabular}

\resizebox{0.48\textwidth}{!}{
\begin{tabular}{|c|cc|}
\hline
                                      & \multicolumn{2}{c|}{Średnia liczba ładowań BSP (SLLB)} \\ \cline{2-3}
\multirow{-2}{*}{}                    & \multicolumn{1}{c|}{\quad\textit{Bez OZE}\quad\quad}       & \multicolumn{1}{c|}{\textit{PVP}}         \\ \hline
Równonoc wiosenna                        & \multicolumn{1}{c|}{$9,03$}                                                & $8,16$                        \\ \hline
Przesilenie letnie                       & \multicolumn{1}{c|}{$9,06$}                                                & $8,1$                        \\ \hline
Równonoc jesienna                        & \multicolumn{1}{c|}{$9,03$}                                                & $8,4$                        \\ \hline
Przesilenie zimowe                       & \multicolumn{1}{c|}{$9,03$}                                                & $8,93$                        \\ \hline
\rowcolor[HTML]{6665CD} 
{\color[HTML]{FFFFFF} Wartość średnia} & \multicolumn{1}{c|}{\cellcolor[HTML]{6665CD}{\color[HTML]{FFFFFF} $9,03$}} & {\color[HTML]{FFFFFF} $8,4$} \\ \hline
\end{tabular}}
\end{table}
\section{Wyniki}
\label{section:results}
Wyniki przeprowadzonych symulacji zostały załączone w~Tab.~\ref{table:energy_characteristics}. Pierwsza tabela przedstawia średnią ilość energii, którą panele fotowoltaiczne mogą zebrać w ciągu całego roku dla pojedynczej stacji BSP, z wyszczególnieniem każdego sezonu. Zgodnie z początkowymi przewidywaniami, największą ilość zasobów, jaką mobilna stacja bazowa jest w stanie uzyskać z promieniowania słonecznego, przypada na przesilenie letnie $\left(572,64\right)$, kiedy to szczytowa wartość procesu produkcji energii jest również najwyższa $\left(91,86\right)$. Kolejno w rankingu uplasowały się wiosenna i jesienna równonoc oraz na końcu przesilenie zimowe. Można zatem zauważyć, że pod względem redukcji energii dostarczanej ze źródeł konwencjonalnych (tj. z akumulatorów ładowanych ze dedykowanych stacji dokowania) przedstawionej w środkowej tabeli, kolejność jest adekwatna do wspomnianych powyżej zależności. Jednakże ze względu na ograniczenia związane z liczbą ogniw fotowoltaicznych oraz ich wydajnością wytwarzania energii, maksymalny uzyskany zysk energetyczny wyniósł $8,71\%$ (przesilenie letnie). Należy również zauważyć, że dla kontrastu podczas przesilenia zimowego ten zysk jest bardzo znikomy $\left(0,27\%\right)$. Wreszcie ostatnia tablica prezentuje średnią liczbę ładowań stacji BSP po rozładowaniu jej systemu bateryjnego. Ze względu na panujące warunki pogodowe, zróżnicowanie tej liczby można zaobserwować nawet wtedy, gdy mobilny punkt dostępowy nie jest zasilany przez generatory OZE. Najwyższą liczbę ładowań dronów odnotowano podczas przesilenia letniego $\left(9,06\right)$, przy czym dla pozostałych dat wartość ta była jednakowa $\left(9,03\right)$. Z drugiej strony, gdy urządzenie BSP jest obsługiwane przez panele fotowoltaiczne, podczas sezonu letniego wymaga ono najmniejszej średniej liczby ładowań baterii $\left(8,1\right)$. Dość podobny zysk odnotowano dla równonocy wiosennej $\left(8,16\right)$, a kolejno za nią znalazła się równonoc jesienna, dla której liczba ładowań baterii dronu (z użyciem paneli solarnych) była równa jej wartości średniej dla pojedynczego dnia w roku $\left(8,4\right)$. Tym samym, ze względu na niemal zerowy wpływ wykorzystania paneli fotowoltaicznych w okresie przesilenia zimowego na charakterystykę bilansu energetycznego sieci, liczba ładowań stacji BSP wykorzystującej OZE zmniejszyła się w tym czasie w najmniejszym stopniu $\left(8,93\right)$ względem przypadku z ich pominięciem $\left(9,03\right)$.

\section{Wnioski}
\label{section:conclusions}
Zaprezentowana w artykule praca podkreśla zalety związane z wykorzystaniem paneli fotowoltaicznych jako generatorów energii w sieciach komórkowych wyposażonych w~urządzenia BSP \textit{na uwięzi} oraz matryce IPR. Dla rozważanego scenariusza systemowego, ze względu na zarówno panujące w Polsce warunki pogodowe, jak i przyjęte konfiguracje stacji BSP oraz generatorów OZE, redukcja zużycia energii z KZE (i wynikające z tego oszczędności finansowe) wynoszą średnio $5,4\%$ rocznie w porównaniu do przypadku, w którym stacje bazowe systemu bezprzewodowego nie są zasilane z OZE. Chociaż odnawialne źródła energii, takie jak panele fotowoltaiczne, charakteryzują się zmiennymi w czasie i zależnymi od klimatu procesami pozyskiwania energii, to poprzez odpowiednie zarządzanie dostępnymi zasobami (radiowymi i energetycznymi) przy pomocy algorytmów zarządzania siecią (jak np. sterowanie ruchem telekomunikacyjnym, alokacja zasobów itp.) oraz dzięki dodatkowemu wyposażeniu, takiemu jak macierze IPR, istnieje możliwość poprawy uzyskanych rezultatów, a nawet zapewnienia autonomii energetycznej sieciom komórkowym bez pogarszania jakości dostarczanych usług mobilnych. Jednakże zarówno implementacja wspomnianych algorytmów, jak i badania ukierunkowane na ewaluację wpływu urządzeń typu IPR na propagację sygnału radiowego będą brane pod uwagę w przyszłych rozważaniach.

\section*{Podziękowania}
\label{section:acknowledgment}

Autorzy pragną podziękować prof. Margot Deruyck z~Uniwersytetu w Gandawie –- IMEC w Belgii za wsparcie niniejszej pracy poprzez udostępnienie programu GRAND. Praca została zrealizowana w ramach projektu nr $2021/43/\text{B}/\text{ST}7/01365$ ufundowanego przez Narodowe Centrum Nauki (NCN) w Polsce.

\end{multicols}
\end{document}